\documentclass[a4paper,11pt]{article}
\usepackage{pos}
\usepackage{graphicx,epsfig,dcolumn,bm,epic,eepic,float}
\usepackage{amsmath}
\usepackage[T1]{fontenc}
\usepackage[utf8]{inputenc}
\usepackage{mathtools} 
\usepackage{pifont}
\usepackage{latexsym}
\usepackage{color}
\usepackage{cancel}
\usepackage{scrextend}
\usepackage{autobreak}
\usepackage{eucal}
\usepackage[bottom]{footmisc}
\usepackage{makeidx,shortvrb,latexsym}
\usepackage{epstopdf}
\usepackage{ragged2e}
\usepackage[numbers,sort&compress]{natbib}
\usepackage{epsfig, axodraw}
\usepackage{graphicx}    
\usepackage{graphics}
\usepackage{afterpage}
\usepackage{array}
\usepackage[flushleft]{threeparttable}
\usepackage{subcaption}
\usepackage{cancel}
\usepackage[mathlines]{lineno}
\usepackage{lettrine}
\usepackage{multirow}
\usepackage{longtable}
\usepackage{enumitem}

\renewcommand{\thefootnote}{\fnsymbol{footnote}}

\newcommand{\nc}{\newcommand}

\newlength{\dhatheight}

\newcommand{\D}[2][]{\ensuremath{\operatorname{d}^{#1}\!{#2}}}
\nc{\beq}{\begin{equation}}  \nc{\eeq}{\end{equation}}
\nc{\bea}{\begin{eqnarray}}  \nc{\eea}{\end{eqnarray}}
\nc{\baa}{\begin{array}}     \nc{\eaa}{\end{array}}
\pdfoutput=1

\DeclareMathAlphabet{\mathcal}{OMS}{cmsy}{m}{n}

\title{}

\title{Probing the Dark Matter EFT with QUEST-DMC: Projected Sensitivities and Attenuation Ceilings} 
\author[]{QUEST-DMC Collaboration}
\author*[a]{N. Darvishi}
 \author[b]{S. Autti}
 \author[c]{L. Bloomfield}
 \author[a]{A. Casey}
 \author[a]{N. Eng}
 \author[c]{P. Franchini}
 \author[b]{R. P. Haley}
 \author[a]{P. J. Heikkinen}
 \author[e]{A. Jennings}
 \author[f]{A. Kemp}
 \author[c,a]{E. Leason}
 \author[c]{J. March-Russell}
 \author[b]{A. Mayer}
 \author[c]{J. Monroe}
 \author[b]{D. M{\"u}nstermann}
 \author[b]{M. T. Noble}
 \author[b]{J. R. Prance}
 \author[a]{X. Rojas}
 \author[b]{T. Salmon}
 \author[a]{J. Saunders}
 \author[d]{J. Smirnov}
 \author[a,c]{R. Smith}
 \author[b]{M. D. Thompson}
 \author[b]{A. Thomson}
 \author[a]{A. Ting}
 \author[b]{V. Tsepelin}
 \author[a]{S. M. West}
 \author[b]{L. Whitehead}
 \author[b]{D. E. Zmeev}

 \affiliation[a]{\textit{Department of Physics, Royal Holloway University of London, Egham, Surrey, TW20 0EX, UK}}
\affiliation[b]{\textit{Department of Physics, Lancaster University, Lancaster, LA1 4YB, UK}}
\affiliation[c]{\textit{Department of Physics, University of Oxford, Keble Road, Oxford, OX1 3RH, UK}}
\affiliation[d]{\textit{Department of Mathematical Sciences, University of Liverpool, Liverpool, L69 7ZL, UK}}
\affiliation[e]{\textit{RIKEN Center for Quantum Computing, RIKEN, Wako, 351-0198, Japan}}
\affiliation[f]{\textit{UKRI STFC Rutherford Appleton Laboratory, Particle Physics Department, Harwell, Didcot OX11 0QX, UK}}

\emailAdd{neda.darvishi@rhul.ac.uk}

\abstract{This proceedings contribution summarises projected constraints from the QUEST-DMC concept, a surface-based direct-detection experiment using superfluid $^3$He operated below the millikelvin regime and instrumented with nanomechanical resonators read out by SQUIDs. The low recoil-energy threshold (down to sub-eV for the SQUID configuration) enables sensitivity to sub-GeV dark matter across a wide set of interaction structures beyond the canonical spin-independent and spin-dependent limits. We present projections in the non-relativistic Effective Field Theory (EFT) framework, scanning the standard set of fourteen Galilean-invariant operators and expressing reach in terms of effective dark matter-nucleon (or dark matter-neutron) cross sections. Because QUEST-DMC operates at the surface, we also account for suppression of the incident flux due to scattering in the atmosphere and Earth, which produces an interaction-dependent sensitivity ceiling at large couplings. Finally, we outline how the non-relativistic results map onto representative relativistic EFT dark matter-nucleon bilinears, enabling a compact interpretation of the projected reach in terms of UV-motivated coupling structures.
 }

\FullConference{}

\begin{document}
\maketitle 

\section{Introduction}
A wide range of astrophysical and cosmological observations indicate that most of the matter content of the Universe is non-baryonic. Direct-detection experiments search for rare energy depositions from dark matter scattering in terrestrial targets. While much of the historical program targeted weak-scale candidates, null results have intensified interest in lighter dark matter, where recoil energies fall well below the keV scale and motivate detectors with ultra-low thresholds.

QUEST-DMC (Quantum Enhanced Superfluid Technologies for Dark Matter and Cosmology) targets the low-mass frontier by using a gram-scale superfluid $^3$He bolometer at microkelvin-scale temperatures. Energy deposition in the fluid produces excitations (with $10^{-7}$ eV superfluid gap) that couple to nanomechanical resonators, and two representative readout configurations are used:
(i) a conventional readout with an effective threshold of 31\,eV (conventional),
(ii) a SQUID-based readout with sub-eV effective threshold 0.51\,eV (SQUID), where low-noise quantum sensor readout enables low-threshold detection ~\cite{QUEST-DMC:2023nug,Autti:2024awr,QUEST-DMC:2025miz,QUEST-DMC:2025qsa,QUEST-DMC:2025ieo}. 

Cosmogenic backgrounds are the primary limiting factor for dark matter searches operated at the surface. To reject these background events from the dark matter search, QUEST-DMC is developing a cosmic-ray muon veto system compatible with operating in a dilution refrigerator. Silicon photomultipliers (SiPMs) have been identified as a candidate photon sensor for this system, and have been recently operated and characterised at $9.4\pm0.2$\,mK inside a dilution refrigerator. Also demonstrated are first proof-of-concept measurements of using a SiPM coupled to scintillator towards identifying high-energy events consistent with cosmic-ray muon signals~\cite{QUEST-DM:2025oep}.

The QUEST-DMC experiment is designed to reach world-leading sensitivity to low-mass dark matter, with particular strength for spin-dependent (SD) dark matter-neutron scattering and competitive reach for spin-independent (SI) dark matter-nucleon scattering, at masses below 0.025\,GeV/$c^2$.
To compare a broad class of dark matter-nucleon interaction structures on equal footing, we adopt the Galilean-invariant Non-Relativistic Effective Field Theory (NREFT) description~\cite{Fan:2010gt,Fitzpatrick:2012ix}. The spin-$1/2$ nature of the $^3$He target provides sensitivity to both SI and SD responses, as well as momentum- and velocity-dependent operator structures, making the NREFT framework particularly well suited. In this approach, the interaction Hamiltonian is written as \begin{equation} \hat{{\cal H}}=\sum_{\tau=0,1}\sum_{i=1}^{15} c_i^\tau {\cal O}_i t^\tau, \label{eq:H_intro} \end{equation} where $\mathcal O_i$ are NREFT operators and $c_i^\tau$ are Wilson coefficients in the isoscalar/isovector basis. The predicted differential recoil rate takes the standard form
\begin{equation}
\frac{\D R}{\D E_R} = \frac{\rho_\chi}{2 \pi m_\chi } \int_{v>v_{\rm min}} \frac{f(\vec{v})}{v} \overline{|\mathcal{M}(v^2, q^2)|^2} \D[3]{v},
\label{eq:rate_intro}
\end{equation}
with $\rho_\chi$ the local dark matter density, $f(\bm{v})$ the lab-frame speed distribution, and $v_{\min}$ the minimum speed required to produce recoil energy $E_R$.  
For surface operation we present projected sensitivities in terms of an operator-dependent \emph{sensitivity floor} (computed assuming an unattenuated halo flux) and an \emph{attenuation ceiling} at large couplings, where scattering in the Earth and atmosphere suppresses the incident flux and limits the maximum cross sections that can be probed~\cite{QUEST-DMC:2025miz,QUEST-DMC:2025qsa}. Atmospheric attenuation becomes particularly relevant for large cross sections, where multiple scatterings modify the velocity distribution and reduce the number of particles capable of producing detectable recoils~\cite{Collar:1992qc,Collar:1993ss,Hasenbalg:1997hs,Kouvaris:2014lpa,Kouvaris:2015laa,Bernabei:2015nia}. At sufficiently large cross sections the Earth becomes effectively opaque, so that the detectable flux is predominantly incident from above after traversing only the atmosphere. The dominant atmospheric targets are $^{14}$N and $^{16}$O (together $>99\%$ by volume); for SD interactions only $^{14}$N contributes, while for SI interactions both $^{14}$N and $^{16}$O are relevant.


\section{NREFT Ingredients for $^3$He and Mapping to Relativistic EFT Couplings}
In the NREFT description, the operator basis is constructed from the Hermitian building blocks
$$
1_{\chi},\quad 1_{N},\quad i\,\frac{\vec q}{m_N},\quad \vec v^{\,\perp},\quad \vec S_{\chi},\quad \vec S_{N},
$$
where $1_{\chi}$ and $ 1_{N}$ are identity operators acting on the dark matter and the target, respectively. The transverse relative velocity of the incoming dark matter particle and the target is defined as $\vec{v}^{\perp}~\equiv \vec{v} + \vec{q}/2\mu_{\chi N}$, where $\vec{v}$ is the dark matter velocity in the lab frame, $\vec{q}$ is the momentum transfer, $m_N$ is the nuclear mass and $\mu_{\chi N} = m_{\chi} m_N/(m_{\chi} + m_N)$ is the dark matter-nucleus reduced mass. With these ingredients, the standard Galilean-invariant operator basis (omitting $\mathcal O_2$) is~\cite{Fan:2010gt,Fitzpatrick:2012ix}:
\begin{equation}
\small
\setlength{\jot}{1pt}
\begin{alignedat}{2}
\mathcal O_{1}  &= 1_\chi 1_N, 
&\hspace{1.6cm}\mathcal O_{3}  &= i\,\vec S_{N}\!\cdot\!\Big(\frac{\vec q}{m_N}\times \vec v^{\perp}\Big),\\
\mathcal O_{4}  &= \vec S_{\chi}\!\cdot\!\vec S_{N}, 
&\hspace{1.6cm}\mathcal O_{5}  &= i\,\vec S_{\chi}\!\cdot\!\Big(\frac{\vec q}{m_N}\times \vec v^{\perp}\Big),\\
\mathcal O_{6}  &= \Big(\vec S_{\chi}\!\cdot\!\frac{\vec q}{m_N}\Big)\Big(\vec S_{N}\!\cdot\!\frac{\vec q}{m_N}\Big), 
&\hspace{1.6cm}\mathcal O_{7}  &= \vec S_{N}\!\cdot\!\vec v^{\perp},\\
\mathcal O_{8}  &= \vec S_{\chi}\!\cdot\!\vec v^{\perp}, 
&\hspace{1.6cm}\mathcal O_{9}  &= i\,\vec S_{\chi}\!\cdot\!\Big(\vec S_{N}\times \frac{\vec q}{m_N}\Big),\\
\mathcal O_{10} &= i\,\vec S_{N}\!\cdot\!\frac{\vec q}{m_N}, 
&\hspace{1.6cm}\mathcal O_{11} &= i\,\vec S_{\chi}\!\cdot\!\frac{\vec q}{m_N},\\
\mathcal O_{12} &= \vec S_{\chi}\!\cdot\!\Big(\vec S_{N}\times \vec v^{\perp}\Big), 
&\hspace{1.6cm}\mathcal O_{13} &= i\,\Big(\vec S_{\chi}\!\cdot\!\vec v^{\perp}\Big)\Big(\vec S_{N}\!\cdot\!\frac{\vec q}{m_N}\Big),\\
\mathcal O_{14} &= i\,\Big(\vec S_{\chi}\!\cdot\!\frac{\vec q}{m_N}\Big)\Big(\vec S_{N}\!\cdot\!\vec v^{\perp}\Big), 
&\hspace{1.6cm}\mathcal O_{15} &= -\Big(\vec S_{\chi}\!\cdot\!\frac{\vec q}{m_N}\Big)\Big[\Big(\vec S_{N}\times \vec v^{\perp}\Big)\!\cdot\!\frac{\vec q}{m_N}\Big].
\end{alignedat}
\label{operators}
\end{equation}

The squared matrix element can be expressed as a sum over nuclear responses $k$ that characterise how a nucleus responds to dark matter interactions~\cite{Fitzpatrick:2012ix,QUEST-DMC:2025miz},
\begin{align}
\overline{|\mathcal{M}(v^2, q^2)|^2} \equiv & {4 \pi \over 2j_N+1}
\sum_{k} \sum_{{\tau,\tau}^\prime=0,1} R^{\tau \tau^\prime}_k \left( \vec{v}_T^{\perp 2}, {\vec{q}^{\,2} \over m_N^2},\left\{c_i^\tau c_j^{\tau^\prime} \right\} \right)
~S_k^{\tau \tau^\prime}(y)\\
\equiv &{4 \pi \over 2j_N+1} \sum_{k} R_k^{pp}S_k^{pp}+R_k^{nn}S_k^{nn}+2R_k^{np}S_k^{np},
\label{Ptot}
\end{align}
where we rewrite the interaction in the proton–neutron basis ($p,\, n$) instead of the isoscalar/isovector basis, and apply the corresponding decomposition consistently to both the dark matter and nuclear response functions. In the general case, the dark matter and nuclear response functions  relevant to $pp$, $nn$, and $np$ interactions can be expressed in terms of the isospin components as
\begin{align}
R^{pp/nn} &= \dfrac{1}{4} \left( R^{00} + R^{11} \pm R^{01} \pm R^{10} \right), \qquad
R^{np}= \dfrac{1}{4} \left( R^{00} - R^{11} \right),
\\
S^{pp/nn} &= S^{00} + S^{11} \pm S^{01} \pm S^{10}, \qquad
S^{np}= S^{00} - S^{11},
\end{align}
with the upper signs corresponding to $pp$ and the lower signs to $nn$. For $^3$He (spin-$1/2$), the relevant nuclear responses entering Eq.~\eqref{Ptot} are the SI $M$ response and the SD responses $\Sigma^\prime$ (transverse) and $\Sigma^{\prime\prime}$ (longitudinal). We evaluate the corresponding dark matter response functions $R_k^{\tau\tau^\prime}$ and nuclear response functions $S_k^{\tau\tau^\prime}(y)$ using the standard NREFT formalism of Ref.~\cite{Fitzpatrick:2012ix} together with the $^3$He response functions from Ref.~\cite{Catena:2015uha}, as implemented in the QUEST-DMC analysis of Ref.~\cite{QUEST-DMC:2025miz}. In the SD case, the $pp$ and $np$ contributions vanish, leaving only the $nn$ response, consistent with the nuclear spin being predominantly carried by the unpaired neutron. In the atmospheric attenuation calculation, additional nuclear responses beyond $M$, $\Sigma^\prime$, and $\Sigma^{\prime\prime}$ contribute for $^{14}$N and $^{16}$O, including $\Phi^{\prime\prime}$, $\tilde\Phi^\prime$, $\Delta$, and interference terms such as $M\Phi^{\prime\prime}$ and $\Delta\Sigma^\prime$; we include the corresponding $R_k^{\tau\tau^\prime}$ and $S_k^{\tau\tau^\prime}$ consistently in the propagation and scattering calculations.

Although interference between multiple non-relativistic operators can occur, we adopt the standard single-operator benchmark in which one coupling dominates at a time. In this limit, all but one coefficient $c_i^{p,n}$ is set to zero, with the basis change $c_i^p=\tfrac{1}{2}(c_i^0+c_i^1)$ and $c_i^n=\tfrac{1}{2}(c_i^0-c_i^1)$. These projections are not intended to represent UV-complete models, but rather provide a transparent benchmark for comparing detector response across different interaction structures.

At low momentum transfer, dark matter interactions may be described either by a relativistic EFT in terms of quark/gluon/photon operators or by the nucleon-level, Galilean-invariant NREFT used here. In general, a single relativistic bilinear matches onto a fixed linear combination of non-relativistic operators. Following Ref.~\cite{Anand:2013yka,QUEST-DMC:2025miz}, Table~\ref{tab:DM_mapping_all} summarises the mapping between representative relativistic dark matter-nucleon bilinears and the corresponding NREFT operator, enabling a compact interpretation of the projected reach in terms of UV-motivated coupling structures (see Ref.~\cite{Bishara:2017pfq} for a quark/gluon-level perspective).


The NREFT basis is the natural language for recoil spectra, while many UV-motivated models are more naturally expressed in terms of relativistic dark matter-nucleon bilinears such as $\bar\chi\chi$, $\bar\chi\gamma^5\chi$, $\bar\chi\gamma^\mu\gamma^5\chi$, and tensor currents, paired with the corresponding nucleon bilinears. In the non-relativistic limit, a given relativistic interaction generally can be reduced to a linear combination of NREFT operators, with coefficients determined by kinematics and mass ratios. Table~\ref{tab:DM_mapping_all} summarises this correspondence in terms of Dirac bilinears $\bar{\chi}\Gamma\chi$ and $\bar{N}\Gamma'N$ (with $\Gamma,\Gamma'\in\{1,\,\gamma^5,\,\gamma^\mu\gamma^5,\, i\sigma^{\mu\nu}q_\nu\}$), and provides a compact interpretation of the projected reach in terms of UV-motivated coupling structures. For example, an axialvector-axialvector interaction maps to a pure $\mathcal{O}_4$ contribution (up to an overall normalisation convention, e.g.\ $-4\,\mathcal{O}_4$ in Ref.~\cite{Anand:2013yka}), corresponding to the familiar SD structure. For readability, Table~\ref{tab:DM_mapping_all} is organised into four blocks according to the nucleon current: (I) scalar ($\bar NN$), (II) axialvector ($\bar N\gamma^\mu\gamma^5 N$), (III) pseudoscalar ($\bar N\gamma^5 N$), and (IV) tensor. The scalar-current block predominantly maps onto SI-type operator content, while the axialvector and pseudoscalar blocks primarily generate SD structures (often with characteristic momentum suppression); tensor currents can yield mixed operator combinations. Momentum insertions are displayed explicitly through the associated $q$- and $K$-dependent prefactors and are normalised by the reference scale $M=0.938$~GeV.

\begin{table}[t]
\centering
\scriptsize
\setlength{\tabcolsep}{3pt}
\renewcommand{\arraystretch}{1.2}

\begin{tabular}{p{0.68\textwidth} p{0.28\textwidth}}
\hline\hline
\textbf{Relativistic operator (spinor-bilinear labelling)} &
\textbf{Factor $\times \mathcal{O}_i$} \\
\hline\hline

\multicolumn{2}{l}{\textbf{(I) Nucleon scalar current: $\bar{N}N$}}\\
\hline
$(\bar{\chi}\chi)(\bar{N}N)$ \newline \textit{Scalar--Scalar} &
$\mathcal{O}_1$ \;[and $4\dfrac{m_\chi m_N}{M^2}\mathcal{O}_1$] \\ \hline

$(\bar{\chi} i\sigma^{\mu\nu}\dfrac{q_\nu}{M}\chi)\,\dfrac{K_\mu}{M}(\bar{N}N)$ \newline \textit{Tensor--Scalar} &
$\displaystyle \frac{1}{M^2}\!\left(\frac{m_N}{m_\chi}\,\vec{q}^{\,2}\mathcal{O}_1-4m_N^2\mathcal{O}_5\right)$ \\ \hline

$(\bar{\chi}\gamma^\mu\gamma^5\chi)\,\dfrac{K_\mu}{M}(\bar{N}N)$ \newline \textit{Axialvector--Scalar} &
$\displaystyle 4\frac{m_N}{M}\mathcal{O}_8$ \\ \hline

$i(\bar{\chi}\gamma^5\chi)(\bar{N}N)$ \newline \textit{Pseudoscalar--Scalar} &
$-\dfrac{m_N}{m_\chi}\mathcal{O}_{11}$ \;[and $-4\dfrac{m_N^2}{M^2}\mathcal{O}_{11}$] \\
\hline\hline

\multicolumn{2}{l}{\textbf{(II) Nucleon axialvector current: $\bar{N}\gamma^\mu\gamma^5N$}}\\
\hline
$(\bar{\chi}\gamma^\mu\gamma^5\chi)(\bar{N}\gamma_\mu\gamma^5N)$ \newline \textit{Axialvector--Axialvector} &
$-4\,\mathcal{O}_4$ \\ \hline

$\dfrac{P^\mu}{M}(\bar{\chi}\chi)(\bar{N}\gamma_\mu\gamma^5N)$ \newline \textit{Scalar--Axialvector} &
$-4\,\dfrac{m_\chi}{M}\mathcal{O}_7$ \\ \hline

$(\bar{\chi} i\sigma^{\mu\nu}\dfrac{q_\nu}{M}\chi)(\bar{N}\gamma_\mu\gamma^5N)$ \newline \textit{Tensor--Axialvector} &
$-4\,\dfrac{m_N}{M}\mathcal{O}_9$ \\ \hline

$i\dfrac{P^\mu}{M}(\bar{\chi}\gamma^5\chi)(\bar{N}\gamma_\mu\gamma^5N)$ \newline \textit{Pseudoscalar--Axialvector} &
$4\,\dfrac{m_N}{M}\mathcal{O}_{14}$ \\
\hline\hline

\multicolumn{2}{l}{\textbf{(III) Nucleon pseudoscalar current: $\bar{N}\gamma^5N$}}\\
\hline
$i(\bar{\chi}\chi)(\bar{N}\gamma^5N)$ \newline \textit{Scalar--Pseudoscalar} &
$\mathcal{O}_{10}$ \;[and $4\dfrac{m_\chi m_N}{M^2}\mathcal{O}_{10}$] \\ \hline

$(\bar{\chi}\gamma^5\chi)(\bar{N}\gamma^5N)$ \newline \textit{Pseudoscalar--Pseudoscalar} &
$-\dfrac{m_N}{m_\chi}\mathcal{O}_6$ \;[and $-4\dfrac{m_N^2}{M^2}\mathcal{O}_6$] \\ \hline

$i(\bar{\chi}\gamma^\mu\gamma^5\chi)\,\dfrac{K_\mu}{M}(\bar{N}\gamma^5N)$ \newline \textit{Axialvector--Pseudoscalar} &
$4\dfrac{m_N}{M}\mathcal{O}_{13}$ \\ \hline

$i(\bar{\chi} i\sigma^{\mu\nu}\dfrac{q_\nu}{M}\chi)\,\dfrac{K_\mu}{M}(\bar{N}\gamma^5N)$ \newline \textit{Tensor--Pseudoscalar} &
$\displaystyle
\begin{aligned}
&\frac{4\vec{q}^{\,2}}{M^2}\!\left(\frac{m_N}{4m_\chi}\mathcal{O}_{10}+\mathcal{O}_{12}\right)+\frac{4m_N^2}{M^2}\mathcal{O}_{15}
\end{aligned}$ \\
\hline\hline

\multicolumn{2}{l}{\textbf{(IV) Nucleon tensor current: $\bar{N}i\sigma_{\mu\alpha}\dfrac{q^\alpha}{M}N$}}\\
\hline
$\dfrac{P^\mu}{M}(\bar{\chi}\chi)\left(\bar{N}i\sigma_{\mu\alpha}\dfrac{q^\alpha}{M}N\right)$ \newline \textit{Scalar--Tensor} &
$\displaystyle \frac{m_\chi m_N}{M^2}\!\left(-\frac{\vec{q}^{\,2}}{m_N^2}\mathcal{O}_1+4\mathcal{O}_3\right)$ \\ \hline

$i\dfrac{P^\mu}{M}(\bar{\chi}\gamma^5\chi)\left(\bar{N}i\sigma_{\mu\alpha}\dfrac{q^\alpha}{M}N\right)$ \newline \textit{Pseudoscalar--Tensor} &
$\displaystyle \frac{1}{M^2}\!\left(\vec{q}^{\,2}\mathcal{O}_{11}+4m_N^2\mathcal{O}_{15}\right)$ \\ \hline

$(\bar{\chi}\gamma^\mu\gamma^5\chi)\left(\bar{N}i\sigma_{\mu\alpha}\dfrac{q^\alpha}{M}N\right)$ \newline \textit{Axialvector--Tensor} &
$4\dfrac{m_N}{M}\mathcal{O}_9$ \\ \hline

$(\bar{\chi} i\sigma^{\mu\nu}\dfrac{q_\nu}{M}\chi)\left(\bar{N}i\sigma_{\mu\alpha}\dfrac{q^\alpha}{M}N\right)$ \newline \textit{Tensor--Tensor} &
$\displaystyle \frac{4}{M^2}\!\left(\vec{q}^{\,2}\mathcal{O}_4-m_N^2\mathcal{O}_6\right)$ \\
\hline\hline
\end{tabular}

\caption{Mapping between representative relativistic dark matter-nucleon bilinear structures and the corresponding non-relativistic EFT operators, following the conventions of Ref.~\cite{Anand:2013yka}. Here $M=0.938$~GeV is the reference mass used to normalise momentum insertions; bracketed terms indicate additional momentum-suppressed contributions.}
\label{tab:DM_mapping_all}
\end{table}

To obtain projected exclusion sensitivities on the full set of relativistic dark matter-nucleon couplings, we construct the complete non-relativistic operator combinations corresponding to each relativistic bilinear structure, following the standard mappings from Lorentz-invariant currents to Galilean-invariant, as summarised in Table~\ref{tab:DM_mapping_all} (Parts I--IV). These combinations typically involve multiple operators with distinct momentum and velocity dependencies and are incorporated directly into the differential recoil rate calculation. Two sets of rates are computed, one assuming unattenuated dark matter flux to determine the experiment's projected sensitivity floor, and another including atmospheric attenuation, which limits the flux of high-cross-section dark matter and defines the sensitivity ceiling. The resulting spectra are then propagated through the full profile likelihood ratio analysis pipeline, which accounts for detector response modelling, energy threshold effects, and expected backgrounds. This procedure preserves the full kinematic structure of each interaction and ensures that spectral features, such as recoil suppression from velocity- or momentum-dependent operators, and distortions arising from broad kinematic distributions, are accurately captured. The inclusion of background expectations enables realistic limit setting under finite exposure, particularly for suppressed interactions near the threshold. Consequently, the projected limits on EFT couplings presented here reflect the detector-level phenomenology of each interaction type, enabling direct comparison across operator classes and consistent interpretation within UV-complete models.

The treatment of dark matter attenuation here assumes the straight-line path (SLP) approximation, in which dark matter particles are assumed to travel directly through the atmosphere, continuously losing energy, while never deviating from the SLP. This assumption is computationally efficient and provides reasonable estimates for heavier dark matter particles, which undergo minimal deflection due to their large mass-to-target mass ratio~\cite{Kouvaris:2015laa,Kavanagh:2017cru}.
For sub-GeV dark matter, a more accurate description is obtained using a diffusion framework which accounts for the effects of multiple random scatterings. In this description, dark matter particles experience both energy loss and angular diffusion, modifying the velocity spectrum at the detector~\cite{Cappiello:2023hza}. Although the diffusion framework offers a more rigorous treatment, the difference in projected sensitivity limits compared to the SLP approximation is approximately a factor of two. For a detailed analysis, see Ref.~\cite{QUEST-DMC:2025qsa}, where we show the ceiling for SI and SD interaction corresponding to the operators $\mathcal{O}_{1}$ and $\mathcal{O}_{4}$, respectively, calculating in both the SLP and diffusive frameworks. The SLP approximation provides a sufficiently accurate sensitivity ceiling for our purpose here, though we note that diffusion effects become important when precision constraints are required, particularly at low masses. 
\section{Projected NREFT Sensitivity, Attenuation Ceilings, and Relativistic Interpretation}
\label{sec:results}

To project sensitivities for various interaction coefficients, we incorporate the QUEST-DMC detector response model described in Ref.~\cite{QUEST-DMC:2023nug} into the expected event rate for each operator. This includes uncertainties from readout noise, fluctuations in quasiparticle production, and shot noise, all bounded by the detector's threshold energy. As in Refs.~\cite{QUEST-DMC:2023nug,QUEST-DMC:2025ieo}, an energy threshold of 31~eV is assumed for the conventional readout with a cold transformer, and 0.51~eV for the SQUID-based readout.
The assumed exposure here is $4.9$~g$\cdot$day, corresponding to five 0.03~g $^3$He cells operated over a six-month cycle.
Because QUEST-DMC is located at the surface, we present sensitivities in terms of both a \emph{sensitivity floor} (computed assuming an unattenuated halo flux) and an \emph{attenuation ceiling} (including the suppression of the incident dark matter flux due to scattering in the atmosphere and Earth for sufficiently large interaction strengths). The ceiling is interaction-dependent and becomes particularly relevant for large cross sections, where multiple scattering events modify the velocity distribution and reduce the population of dark matter particles capable of producing detectable recoils.

\begin{figure*}[t]
\includegraphics[width=0.495\textwidth]{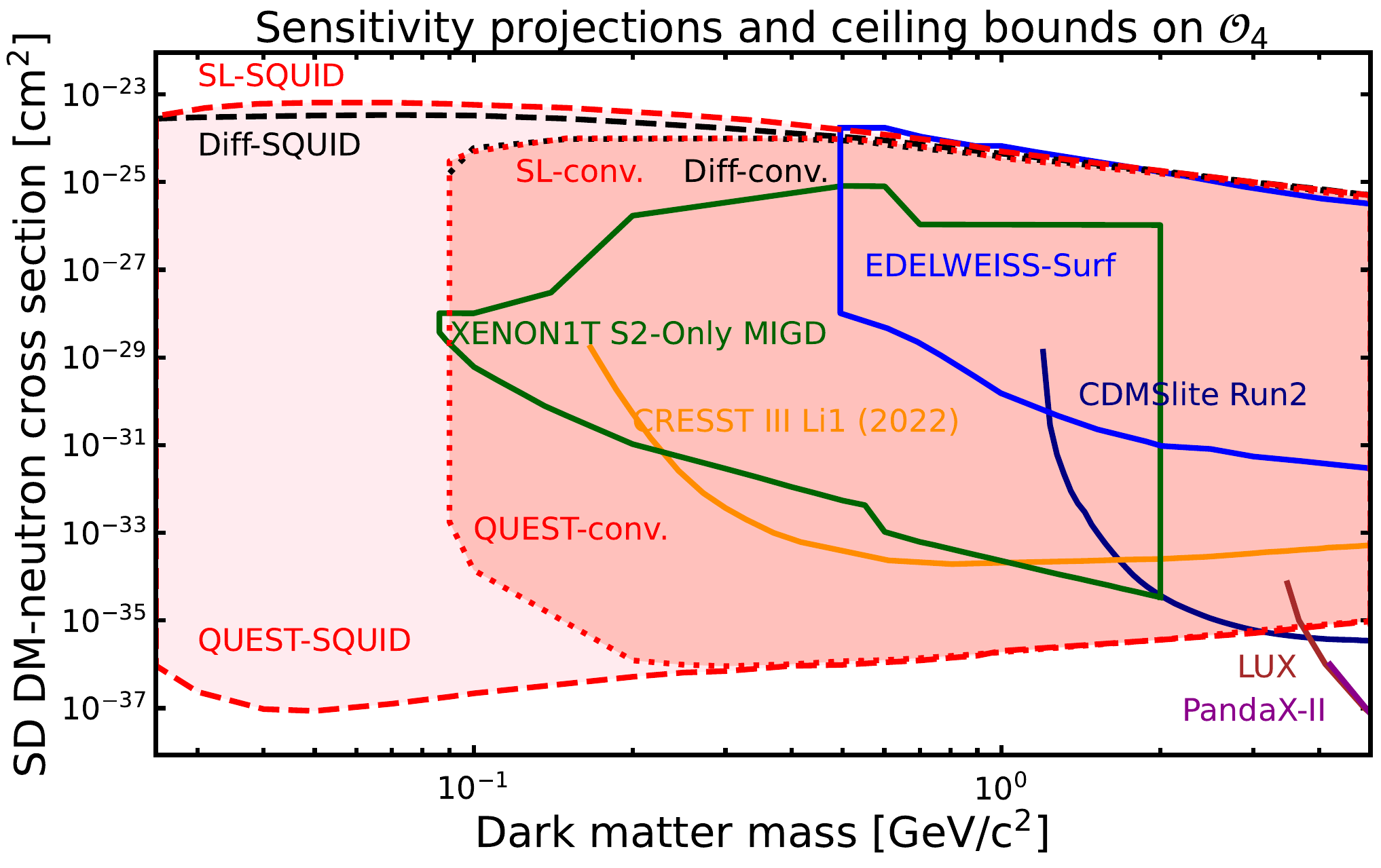}
\includegraphics[width=0.495\textwidth]{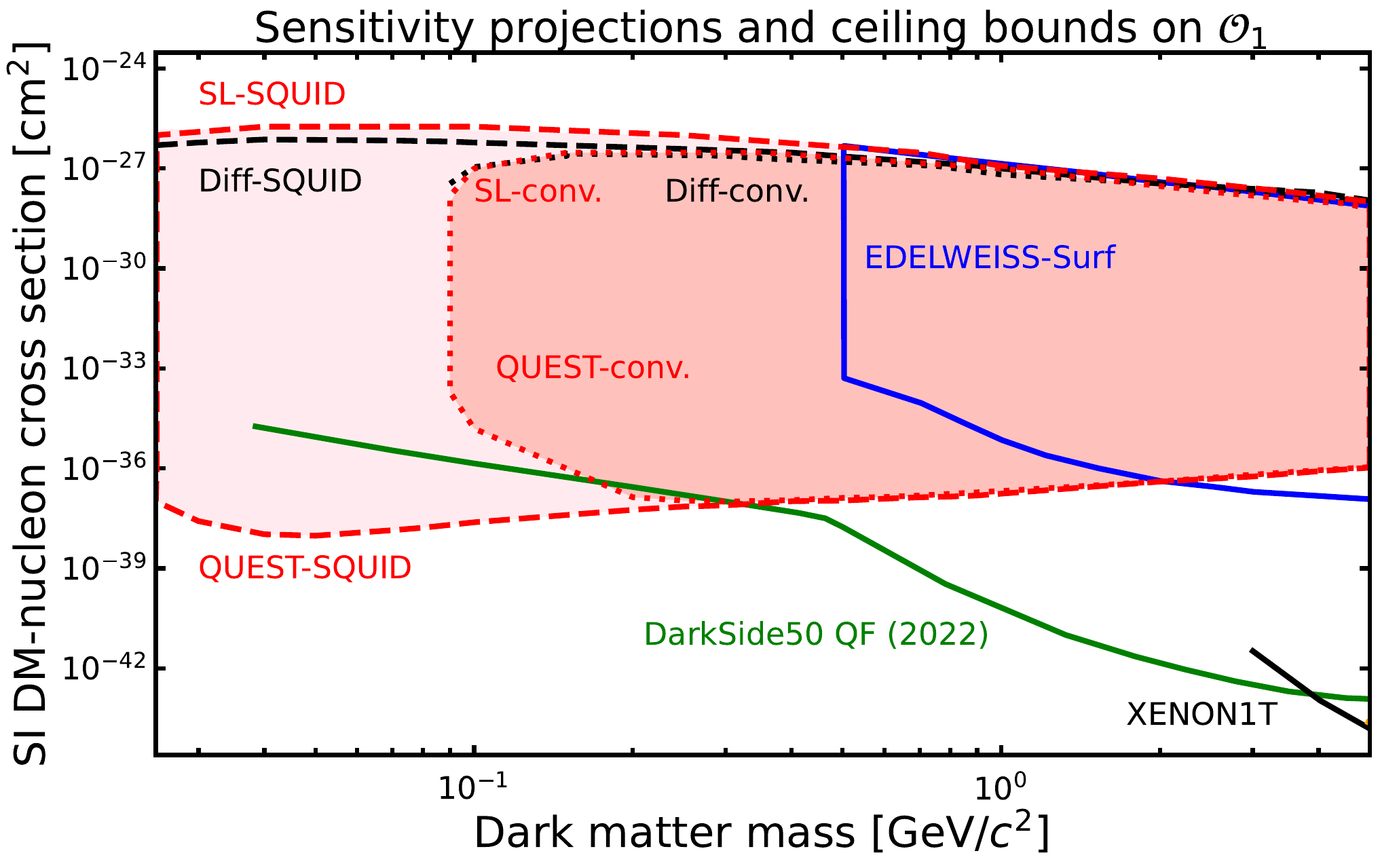}
\caption{The QUEST-DMC 90\%~C.L. limits on the cross-section for velocity- and momentum-independent operators $\mathcal{O}_4$ (SD, left) and $\mathcal{O}_1$ (SI, right), compared to existing limits from Xenon 1T S2-only MIGD~\cite{XENON:2019zpr}, CRESST III (LiAlO$_2$)~\cite{CRESST_2022}, LUX~\cite{LUXSD_2016}, CDMSlite~\cite{CDMSLite_2018}, PandaX-II~\cite{PandaX-II:2018woa} and EDELWEISS~\cite{EDELWEISS:2019vjv} for SD, and from DarkSide-50~\cite{DarkSide-50:2022qzh}, XENON1T~\cite{XENON:2018voc}, and EDELWEISS~\cite{EDELWEISS:2019vjv} for SI. The upper limit sensitivity is shown for the straight-line (SL) path and diffusive (Diff) propagation treatments, depicted in red and black, respectively. Dashed (dotted) lines correspond to SQUID-based (conventional) readout.}
\label{fig:Ops_limits1}
\end{figure*}

The benchmark results for canonical SI and SD interactions are represented by operators $\mathcal{O}_1$ and $\mathcal{O}_4$.
For the SD operator $\mathcal{O}_4$, the SQUID readout yields sensitivity to exclude cross sections as low as $6.5 \times 10^{-24}\,{\rm cm}^2$ under the straight-line path model and $3.3 \times 10^{-24}\,{\rm cm}^2$ with the diffusion framework, within the $0.04$-$0.07$~GeV/$c^2$ mass range.
By contrast, with the conventional cold transformer readout, the sensitivity floor is $1 \times 10^{-24}\,{\rm cm}^2$, covering a higher mass range of $0.1$-$0.55$~GeV/$c^2$ for both propagation models.
For SI interactions via $\mathcal{O}_1$, QUEST-DMC extends reach into the sub-GeV regime; the projected sensitivity reaches down to $\sim 7.5\times 10^{-27}\,{\rm cm}^2$ over $0.025$-$5$~GeV/$c^2$ (SQUID-based configuration). At large cross sections, attenuation of the incident flux sets an interaction-dependent ceiling that limits the accessible parameter space, with differences between the SLP and diffusive treatments typically at the level of a factor of two.

\begin{figure*}[t]
\centering
\includegraphics[width=0.62\textwidth]{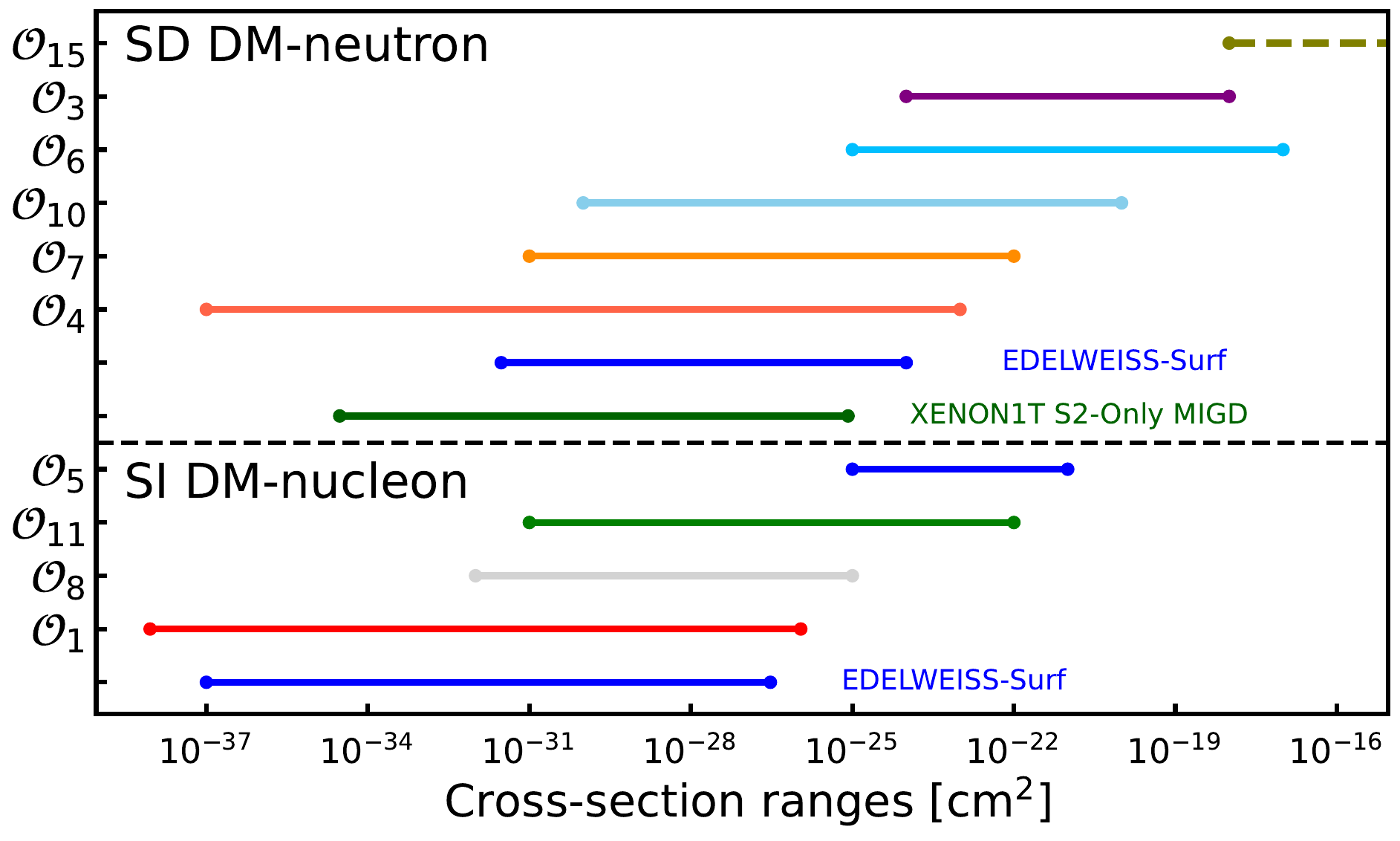}
\caption{Overview of the collective cross-section range probed for the analysed NREFT operators in the SQUID-based QUEST-DMC configuration, shown as the span between the lowest projected sensitivity (floor) and the highest excluded value set by attenuation (ceiling). The summary illustrates the breadth of EFT parameter space covered by QUEST-DMC in the low-mass regime, and enables compact comparison to representative existing limits (e.g.\ Xenon 1T S2-only MIGD~\cite{XENON:2019zpr} and EDELWEISS~\cite{EDELWEISS:2019vjv}, depending on interaction type).}
\label{fig:Ops_limitsmax_summary}
\end{figure*}

Beyond these fundamental interactions, additional EFT operators probe different aspects of dark matter-nucleon interactions, including explicit momentum transfer and transverse-velocity dependence. Representative SI operators such as the velocity-dependent $\mathcal{O}_8$ and momentum-dependent $\mathcal{O}_{11}$ yield cross-section constraints spanning $\sim [10^{-32} - 10^{-26}]\,{\rm cm}^2$ and $\sim [10^{-31} - 10^{-22}]\,{\rm cm}^2$, respectively, while the largest SI cross sections that remain testable before attenuation saturates the sensitivity are associated with the combined velocity- and momentum-dependent operator $\mathcal{O}_5$, which lies within $[10^{-25} - 10^{-20}]\,{\rm cm}^2$ (see Ref.~\cite{QUEST-DMC:2025miz}; summarised in Fig.~\ref{fig:Ops_limitsmax_summary}). For SD interactions, the velocity-dependent operator $\mathcal{O}_7$ and the momentum-dependent operator $\mathcal{O}_{10}$ span projected exclusion sensitivities within $[10^{-31} - 10^{-22}]\,{\rm cm}^2$ and $[10^{-29} - 10^{-20}]\,{\rm cm}^2$, respectively. Simultaneous velocity- and momentum-dependent interactions are captured by operators such as $\mathcal{O}_3$ and $\mathcal{O}_6$, whose lower-limit scale is larger than that of $\mathcal{O}_5$ due to stronger kinematic suppression. Across operators, the SQUID-based readout consistently achieves the strongest projected reach.
The hierarchy of projected event rates across operators is driven by the explicit momentum- and transverse-velocity dependence of the non-relativistic amplitudes, and by how these factors enter the halo integral in Eq.~\eqref{eq:rate_intro}. Operators $\mathcal{O}_1$ and $\mathcal{O}_4$ are unsuppressed in $q$ and $v^\perp$, so they populate the recoil spectrum most efficiently at fixed coupling and therefore yield the deepest sensitivity floors. Operators with a single power of $v^\perp$ or $q$ in the amplitude are kinematically suppressed at low recoil energy and for the typical halo speeds, pushing the sensitivity floor upward. In our point-like normalisation, the fixed-speed angular averages imply a characteristic scaling of the nucleon-level cross section with even powers of the dark matter speed: $\sigma_{\chi n}^{c_i}(v)\propto v^{2}$ for $\mathcal{O}_{7,8,9,10,11,12}$, $\propto v^{4}$ for $\mathcal{O}_{3,5,6,13,14}$, and $\propto v^{6}$ for $\mathcal{O}_{15}$ (with $\mathcal{O}_{1}$ and $\mathcal{O}_{4}$ velocity independent). Since the rate weights the integrand by $f(\bm{v})/v$, higher powers of $v$ increasingly emphasise the high-speed tail of the distribution, while additional powers of $q$ preferentially shift spectral weight toward higher recoils. As a result, mixed $q$--$v^\perp$ operators such as $\mathcal{O}_3$ and $\mathcal{O}_6$, and especially $\mathcal{O}_{15}$, are the most penalised at low mass and near-threshold energies, yielding weaker projected reach for the same coupling.
Atmospheric attenuation further amplifies this ordering for surface operation: by preferentially degrading or removing the fastest dark matter particles, it most strongly impacts operators whose signal rate is dominated by the high-velocity tail (notably those with $\sigma_{\chi n}^{c_i}(v)\propto v^{4}$ or $v^{6}$), thereby setting an operator-dependent sensitivity ceiling at large couplings.

Degeneracies arise between operators that produce similar recoil spectra, resulting in nearly indistinguishable signals within the detector energy sensitivity range. In particular, operators $\mathcal{O}_{13}$ and $\mathcal{O}_{14}$ yield almost identical responses, while pairs such as $\mathcal{O}_9$ and $\mathcal{O}_{10}$, and $\mathcal{O}_7$ and $\mathcal{O}_{12}$, differ only by a small normalisation factor. These near-degeneracies motivate emphasising operator families and kinematic scaling classes rather than treating each operator as fully distinct in spectral shape.

\begin{figure}[t]
\centering
\includegraphics[width=0.6\textwidth]{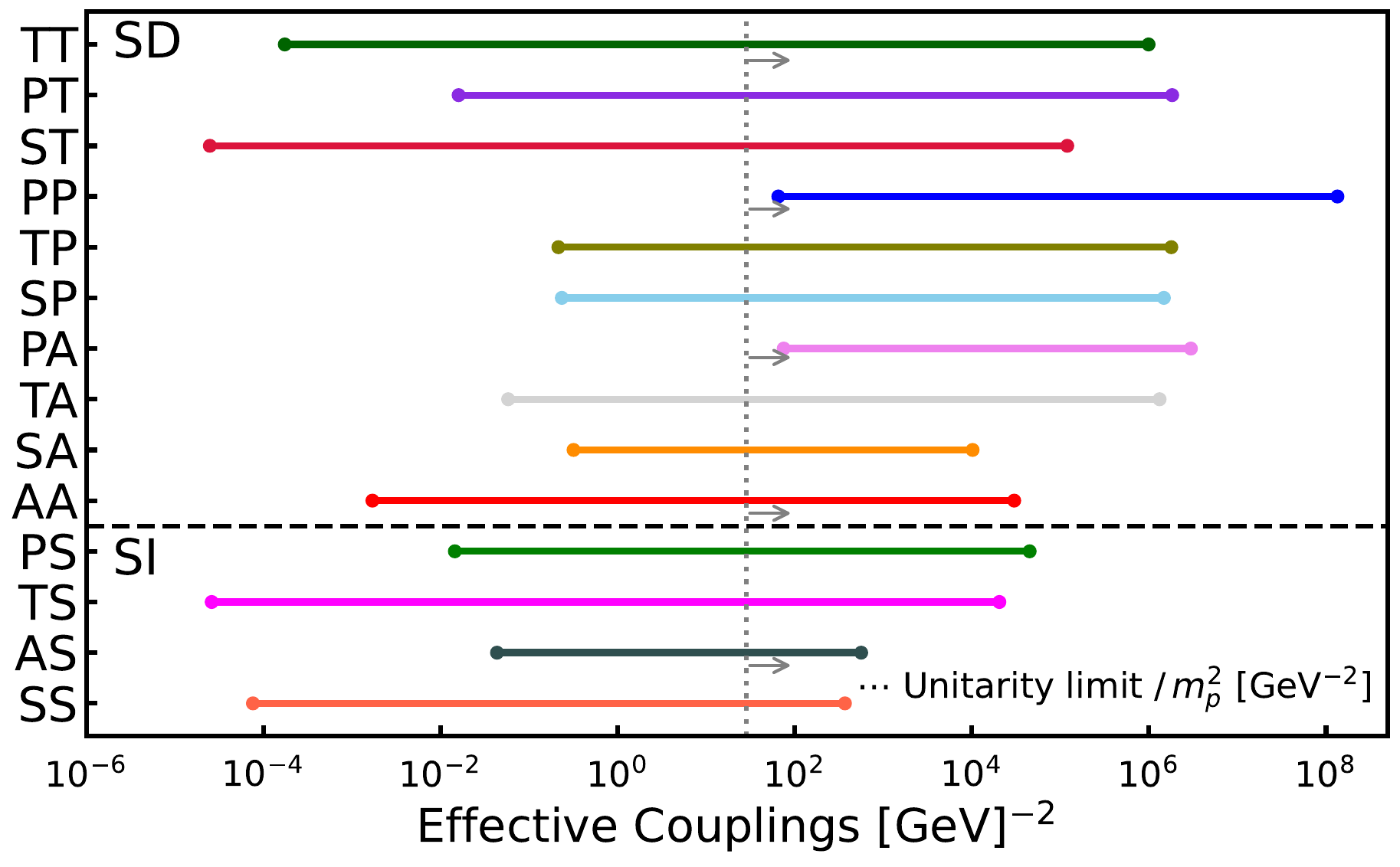}
\caption{
Effective coupling ranges for dark matter-nucleon and -neutron interactions, grouped by their underlying relativistic bilinear types (SS, TS, AA, etc.) as defined by the relativistic to non-relativistic mapping in Table~\ref{tab:DM_mapping_all}. The couplings represent rescaled Wilson coefficients derived from the QUEST-DMC SQUID-based sensitivity projections. The vertical dashed line with right-pointing arrows indicates the region above the unitarity limit.
}
\label{fig:summary_limits}
\end{figure}

To obtain projected exclusion sensitivities on the full set of relativistic dark matter-nucleon couplings, we construct the complete non-relativistic operator combinations corresponding to each relativistic bilinear structure, following the standard mappings from Lorentz-invariant currents to Galilean-invariant EFT operators, as summarised in Table~\ref{tab:DM_mapping_all}. These combinations typically involve multiple operators with distinct momentum and velocity dependencies and are incorporated directly into the differential recoil rate calculation. Two sets of rates are computed, one assuming unattenuated dark matter flux to determine the experiment's projected sensitivity floor, and another including atmospheric attenuation, which limits the flux of high-cross-section dark matter and defines the sensitivity ceiling. The resulting spectra are then propagated through the full profile likelihood ratio analysis pipeline, which accounts for detector response modelling, energy threshold effects, and expected backgrounds. This procedure preserves the full kinematic structure of each interaction and ensures that spectral features, such as recoil suppression from velocity- or momentum-dependent operators, and distortions arising from broad kinematic distributions, are accurately captured. The inclusion of background expectations enables realistic limit setting under finite exposure, particularly for suppressed interactions near the threshold. Consequently, the projected limits on EFT couplings presented here reflect the detector-level phenomenology of each interaction type, enabling direct comparison across operator classes and consistent interpretation within UV-complete models.

We note that for certain interactions, the corresponding non-relativistic operator scaling is indicated in brackets in the mapping table, reflecting the same spinor structure but with additional momentum-suppressed contributions from four-momentum insertions on both the dark matter and nucleon sides; however, these contributions are omitted from the coupling-summary presentation as they do not represent independent physical operators. The PandaX-II experiment~\cite{PandaX-II:2018woa} has placed lower bounds on effective couplings for WIMP masses above $\sim 5~{\rm GeV}/c^2$, which are not shown here as our analysis focuses on the sub-GeV mass regime.

Notably, the projected sensitivity floors indicate that Lorentz structures involving tensor-scalar and scalar-tensor spinor combinations can lead to more stringent sensitivities than the conventionally dominant scalar-scalar and axialvector-axialvector forms. The tensor-scalar case arises from the combination of $\mathcal{O}_1$ and $\mathcal{O}_5$ (both SI), while the scalar-tensor structure maps onto a mix of $\mathcal{O}_1$ and $\mathcal{O}_3$, incorporating both SI and SD contributions. Across the sub-GeV mass range, tensor-scalar interactions yield stronger projected coupling sensitivities than scalar-scalar for dark matter masses above $0.1$~GeV; for example, at $m_\chi = 1~{\rm GeV}/c^2$, the tensor-scalar limit is $3.0 \times 10^{-5}\,{\rm GeV}^{-2}$, compared to $7.7 \times 10^{-5}\,{\rm GeV}^{-2}$ for scalar-scalar. Scalar-tensor spinor combinations consistently yield tighter limits than axialvector-axialvector couplings, reaching $2.84 \times 10^{-5}\,{\rm GeV}^{-2}$ at $1~{\rm GeV}/c^2$ versus $1.83 \times 10^{-3}\,{\rm GeV}^{-2}$. This difference becomes more significant at higher masses. Among all interaction types, the pseudoscalar-pseudoscalar structure, corresponding dominantly to the non-relativistic operator $\mathcal{O}_6$, exhibits one of the highest sensitivity ceilings, consistent with the sharp cross-section growth at low mass. Fig.~\ref{fig:summary_limits} summarises the constrained effective coupling ranges across interaction classes grouped by relativistic bilinear combinations. The horizontal bars span the coupling range from the projected exclusion sensitivity floor to the sensitivity ceiling imposed by atmospheric attenuation. Only results from the SQUID-based readout are shown, as they provide the most stringent projected constraints.

The observed variation in coupling sensitivity across operator classes, in some cases approaching two orders of magnitude, highlights the importance of a systematic and comprehensive mapping between relativistic bilinears and non-relativistic EFT operators. Rather than relying solely on conventional benchmarks such as SI or SD interactions, this framework accounts for Lorentz structures that map onto non-relativistic operators with recoil spectra within the energy sensitivity range of a given detector. These mappings help identify which interactions are more likely to evade or dominate experimental sensitivity, depending on their momentum and velocity dependence, thereby enabling a more accurate connection between low-energy phenomenology and UV-complete theoretical models. However, in realistic UV-complete theories, it is uncommon for a single relativistic dark matter-nucleon operator to arise in isolation. Interactions are typically generated at the dark matter-quark/gluon level and, after hadronisation and matching onto nucleon-level degrees of freedom, give rise to combinations of both relativistic and non-relativistic operators. Our analysis treats individual relativistic bilinears as a phenomenologically useful intermediate step between the non-relativistic EFT framework and UV-complete models. A full treatment, beginning with dark matter-quark/gluon EFT operators and their systematic matching onto nucleons (see e.g.\ Ref.~\cite{Bishara:2017pfq}), would more faithfully reflect the operator structure predicted by UV completions; while beyond the scope of this proceedings contribution, it represents a promising direction for future work.

\section{Conclusions}\label{sec:Concl}
In this proceedings contribution we have summarised projected constraints from the QUEST-DMC experiment, following the full analysis of Ref.~\cite{QUEST-DMC:2025miz}. QUEST-DMC is a surface-based direct-detection concept using superfluid $^3$He operated in the microkelvin regime and instrumented with nanomechanical resonators, with emphasis on the SQUID-based readout configuration with sub-eV threshold. Using the detector response and statistical inference framework developed in Ref.~\cite{QUEST-DMC:2023nug}, we presented projected 90\% C.L. exclusion sensitivities for both canonical and non-canonical dark matter interaction hypotheses in the NREFT description, and highlighted the role of atmospheric (and terrestrial) attenuation in setting an interaction-dependent sensitivity ceiling at large couplings.

For the standard benchmarks, $\mathcal{O}_1$ (SI) and $\mathcal{O}_4$ (SD), QUEST-DMC is projected to probe sub-GeV dark matter with sensitivity well below existing constraints in the low-mass regime. For SD interactions through $\mathcal{O}_4$, the SQUID configuration reaches cross sections down to $6.5\times10^{-24}\,\mathrm{cm}^2$ in the straight-line propagation treatment (and $3.3\times10^{-24}\,\mathrm{cm}^2$ with a diffusive treatment) around $m_\chi\simeq 0.04$-$0.07~\mathrm{GeV}/c^2$, while the conventional readout provides complementary coverage at higher masses. For SI interactions through $\mathcal{O}_1$, the projected reach extends to $\sim 7.5\times10^{-27}\,\mathrm{cm}^2$ over masses from $\sim 0.025$ to a few GeV/$c^2$.

Moving beyond the canonical SI/SD benchmarks, QUEST-DMC probes a broad range of the NREFT operator space, including momentum- and velocity-dependent interactions whose recoil spectra can be strongly suppressed near threshold in conventional detectors. The operator-by-operator hierarchy is governed by the kinematic structure of the interaction: operators free of $q$ or $v^\perp$ suppression produce the strongest rates, while mixed and higher-order structures require larger couplings to yield detectable recoils. For surface operation, these projections must be interpreted together with an attenuation ceiling, since sufficiently large cross sections deplete the incident dark matter population through scattering in the atmosphere and Earth. We also noted near-degeneracies among subsets of operators that generate almost indistinguishable recoil spectra within the relevant energy range, motivating an emphasis on kinematic classes and operator families in compact presentations of projected reach.

Finally, we connected the NREFT results to UV-motivated model building by recasting the projections in terms of relativistic dark matter-nucleon bilinear structures, using the standard mapping between Lorentz-invariant currents and Galilean-invariant non-relativistic operators. This bilinear-grouped representation provides a compact interpretation of the projected reach in coupling space and highlights that non-canonical relativistic structures, such as tensor-scalar and scalar-tensor combinations, can yield projected sensitivities that are competitive with or stronger than the conventional scalar-scalar and axialvector-axialvector benchmarks across portions of the sub-GeV mass range. 

Overall, QUEST-DMC is projected to access previously unexplored parameter space for sub-GeV dark matter across a wide range of EFT interaction structures, while surface operation makes attenuation-induced ceilings an essential part of the interpretation at large couplings. The combination of ultra-low threshold sensitivity, a light nuclear target, and a systematic EFT treatment positions QUEST-DMC as a powerful probe of dark matter interactions beyond the traditional WIMP paradigm.
\subsection*{Acknowledgements} 
This work was funded by UKRI EPSRC and STFC (Grants ST/T006773/1, ST/Y004434/1, EP/P024203/1, EP/W015730/1 and EP/W028417/1), as well as the European Union's Horizon 2020 Research and Innovation Programme under Grant Agreement no 824109 (European Microkelvin Platform). S.A. acknowledges financial support from the Jenny and Antti Wihuri Foundation. M.D.T acknowledges financial support from the Royal Academy of Engineering (RF/201819/18/2). J.S. acknowledges support from the UK Research and Innovation Future Leader Fellowship~MR/Y018656/1. A.K. acknowledges support from the UK Research and Innovation Future Leader Fellowship MR/Y019032/1.
\noindent

\bibliographystyle{unsrt}
\bibliography{biblio}

\begin{thebibliography}{10}

\bibitem{QUEST-DMC:2023nug}
S.~Autti et~al.
\newblock {QUEST-DMC superfluid $^3$He detector for sub-GeV dark matter}.
\newblock {\em Eur. Phys. J. C}, 84(3):248, 2024.

\bibitem{Autti:2024awr}
S.~Autti et~al.
\newblock {QUEST-DMC: Background Modelling and Resulting Heat Deposit for a
  Superfluid Helium-3 Bolometer}.
\newblock {\em J. Low Temp. Phys.}, 215(5-6):465--476, 2024.

\bibitem{QUEST-DMC:2025miz}
N.~Darvishi et~al.
\newblock {Dark Matter EFT landscape probed by QUEST-DMC}.
\newblock {\em JCAP}, 10:044, 2025.

\bibitem{QUEST-DMC:2025qsa}
N.~Darvishi et~al.
\newblock {Dark matter attenuation effects: sensitivity ceilings for
  spin-dependent and spin-independent interactions}.
\newblock {\em JCAP}, 04:017, 2025.

\bibitem{QUEST-DMC:2025ieo}
E.~Leason et~al.
\newblock {Development of Superfluid Helium-3 Bolometry Using Nanowire
  Resonators with SQUID Readout for the QUEST-DMC Experiment}.
\newblock {\em J. Low Temp. Phys.}, 222(2):39, 2026.

\bibitem{QUEST-DM:2025oep}
A.~Kemp et~al.
\newblock {Operation of silicon photomultipliers in a dilution refrigerator
  down to 9.4 mK towards a cryogenic cosmic ray muon veto system}.
\newblock arXiv: 2512.16769.

\bibitem{Fan:2010gt}
JiJi Fan, Matthew Reece, and Lian-Tao Wang.
\newblock {Non-relativistic effective theory of dark matter direct detection}.
\newblock {\em JCAP}, 11:042, 2010.

\bibitem{Fitzpatrick:2012ix}
A.~Liam Fitzpatrick, Wick Haxton, Emanuel Katz, Nicholas Lubbers, and Yiming
  Xu.
\newblock {The Effective Field Theory of Dark Matter Direct Detection}.
\newblock {\em JCAP}, 02:004, 2013.

\bibitem{Collar:1992qc}
J.~I. Collar and F.~T. Avignone.
\newblock {Diurnal modulation effects in cold dark matter experiments}.
\newblock {\em Phys. Lett. B}, 275:181--185, 1992.

\bibitem{Collar:1993ss}
J.~I. Collar and F.~T. Avignone, III.
\newblock {The Effect of elastic scattering in the Earth on cold dark matter
  experiments}.
\newblock {\em Phys. Rev. D}, 47:5238--5246, 1993.

\bibitem{Hasenbalg:1997hs}
F.~Hasenbalg et~al.
\newblock {Cold dark matter identification: Diurnal modulation revisited}.
\newblock {\em Phys. Rev. D}, 55:7350--7355, 1997.

\bibitem{Kouvaris:2014lpa}
C.~Kouvaris and I.~M. Shoemaker.
\newblock {Daily modulation as a smoking gun of dark matter with significant
  stopping rate}.
\newblock {\em Phys. Rev. D}, 90:095011, 2014.

\bibitem{Kouvaris:2015laa}
C.~Kouvaris.
\newblock {Earth\textquoteright{}s stopping effect in directional dark matter
  detectors}.
\newblock {\em Phys. Rev. D}, 93(3):035023, 2016.

\bibitem{Bernabei:2015nia}
R.~Bernabei et~al.
\newblock {Investigating Earth shadowing effect with DAMA/LIBRA-phase1}.
\newblock {\em Eur. Phys. J. C}, 75(5):239, 2015.

\bibitem{Catena:2015uha}
Riccardo Catena and Bodo Schwabe.
\newblock {Form factors for dark matter capture by the Sun in effective
  theories}.
\newblock {\em JCAP}, 04:042, 2015.

\bibitem{Anand:2013yka}
Nikhil Anand, A.~Liam Fitzpatrick, and W.~C. Haxton.
\newblock {Weakly interacting massive particle-nucleus elastic scattering
  response}.
\newblock {\em Phys. Rev. C}, 89(6):065501, 2014.

\bibitem{Bishara:2017pfq}
Fady Bishara, Joachim Brod, Benjamin Grinstein, and Jure Zupan.
\newblock {From quarks to nucleons in dark matter direct detection}.
\newblock {\em JHEP}, 11:059, 2017.

\bibitem{Kavanagh:2017cru}
Bradley~J. Kavanagh.
\newblock {Earth scattering of superheavy dark matter: Updated constraints from
  detectors old and new}.
\newblock {\em Phys. Rev. D}, 97(12):123013, 2018.

\bibitem{Cappiello:2023hza}
Christopher~V. Cappiello.
\newblock {Analytic Approach to Light Dark Matter Propagation}.
\newblock {\em Phys. Rev. Lett.}, 130(22):221001, 2023.

\bibitem{XENON:2019zpr}
E.~Aprile et~al.
\newblock {Search for Light Dark Matter Interactions Enhanced by the Migdal
  Effect or Bremsstrahlung in XENON1T}.
\newblock {\em Phys. Rev. Lett.}, 123(24):241803, 2019.

\bibitem{CRESST_2022}
G.~Angloher and others.
\newblock Testing spin-dependent dark matter interactions with lithium
  aluminate targets in {CRESST-III}.
\newblock {\em Phys. Rev. D}, 106:092008, Nov 2022.

\bibitem{LUXSD_2016}
D.{\hspace{0.167em}}S. Akerib et~al.
\newblock Results on the spin-dependent scattering of weakly interacting
  massive particles on nucleons from the {Run 3} data of the {LUX} experiment.
\newblock {\em Phys. Rev. Lett.}, 116(16), Apr 2016.

\bibitem{CDMSLite_2018}
R.~Agnese et~al.
\newblock Low-mass dark matter search with {CDMSlite}.
\newblock {\em Phys. Rev. D}, 97(2), Jan 2018.

\bibitem{PandaX-II:2018woa}
J.~Xia et~al.
\newblock {PandaX-II Constraints on Spin-Dependent WIMP-Nucleon Effective
  Interactions}.
\newblock {\em Phys. Lett. B}, 792:193--198, 2019.

\bibitem{EDELWEISS:2019vjv}
E.~Armengaud et~al.
\newblock {Searching for low-mass dark matter particles with a massive Ge
  bolometer operated above-ground}.
\newblock {\em Phys. Rev. D}, 99(8):082003, 2019.

\bibitem{DarkSide-50:2022qzh}
P.~Agnes et~al.
\newblock {Search for low-mass dark matter WIMPs with 12~ton-day exposure of
  DarkSide-50}.
\newblock {\em Phys. Rev. D}, 107(6):063001, 2023.

\bibitem{XENON:2018voc}
E.~Aprile et~al.
\newblock {Dark Matter Search Results from a One Ton-Year Exposure of XENON1T}.
\newblock {\em Phys. Rev. Lett.}, 121(11):111302, 2018.

\end{thebibliography}

\end{document}